# PieGlyph: An R package for creating axis invariant pie-glyphs for 2d plots

Rishabh Vishwakarma[1], Caroline Brophy[1], and Catherine Hurley[2]

[1]School of Computer Science and Statistics, Trinity College Dublin, Dublin 2, Ireland.

[2]Department of Mathematics and Statistics, Maynooth University, Maynooth, Co. Kildare, Ireland.

**Corresponding Author**

Rishabh Vishwakarma

Email: vishwakr@tcd.ie

Address: School of Computer Science and Statistics, Trinity College Dublin, Ireland

# Summary

Effective visualisations are crucial for understanding and deriving insights from multidimensional data. Glyph-based visualisations encode multiple data variables onto aesthetic attributes of a graphical symbol (Ward, 2008), and additionally, the glyph position may encode two more variables in a 2d plot. A major benefit of glyphs is the ability to offer a compact and direct representation of data observations, thereby facilitating richer pattern recognition. Pie-glyphs are a glyph-based visualisation approach that show different data dimensions as a pie-chart. Pie-charts are best suited for proportional data as the angle/area of the sector (or slice) within the pie-chart corresponds to relative magnitudes of the respective data attributes. Overlaying a traditional 2d plot with pie-chart glyphs (pie-glyphs) can lead to more insightful visualisations. In fact, the use of pie-glyphs overlaid on a map dates back to Minard who in 1858 published a map of France with pie-charts showing the meat type proportions exported by departments to Paris, and pie size representing the total amount of exports (see for example Plate 6 in Friendly and Wainer, 2021). As Friendly and Wainer, 2021 (page 104) says: "Thus the humble pie, when combined with other graphic forms (here a map) can quickly convey a complex story".

The `PieGlyph` package (Vishwakarma, Brophy, and Hurley, 2024) developed for R software (R Core Team, 2023) enables users to overlay any 2d plot with pie-glyphs. Figure 1 shows the additional insights offered by a pie-glyph scatterplot over a traditional scatterplot, using data consisting of ratings and sales figures of selected video games (Kirubi, 2017). In addition to showing the positive relationship between user and critic scores for various games, the pie-glyphs in Figure 1 give information about the relative sales for each game in North America, Europe, Japan, and the rest of the world. It is clear that North America and Europe account for the majority of sales for almost all games. Two pie-glyphs are labelled to illustrate the effects of user preferences on the distribution of game sales. For example, *FIFA 17* has majority sales in Europe while *NBA 2K17* has majority sales in North America. Generally, games with a lower critic score did not sell across multiple regions (expected as users would refrain from purchasing a poorly reviewed game), while those with higher critic scores had sales in multiple regions. Thus, the inclusion of pie-glyphs provides additional insights for this multi-dimensional data.

# Statement of need

`PieGlyph` is developed under the Grammar of Graphics (Wilkinson, 2012) paradigm using the `ggplot2` (Wickham, 2016) plotting framework. Thus, in addition to showing data attributes along the pie-chart slices, the `ggplot2` machinery can be leveraged to facet the plot (i.e., subset the data and show each subset in a different panel) based on additional data attributes; or map additional data attributes onto aesthetics such as the size, colour, or style of the pie-glyphs to jointly visualise even more data dimensions. For example, in the pie-glyph size could encode the total sales for each game, or the plot could be faceted based on game genre, allowing for the visualization of user and critics scores across different genres while also showing regional preferences for each genre.

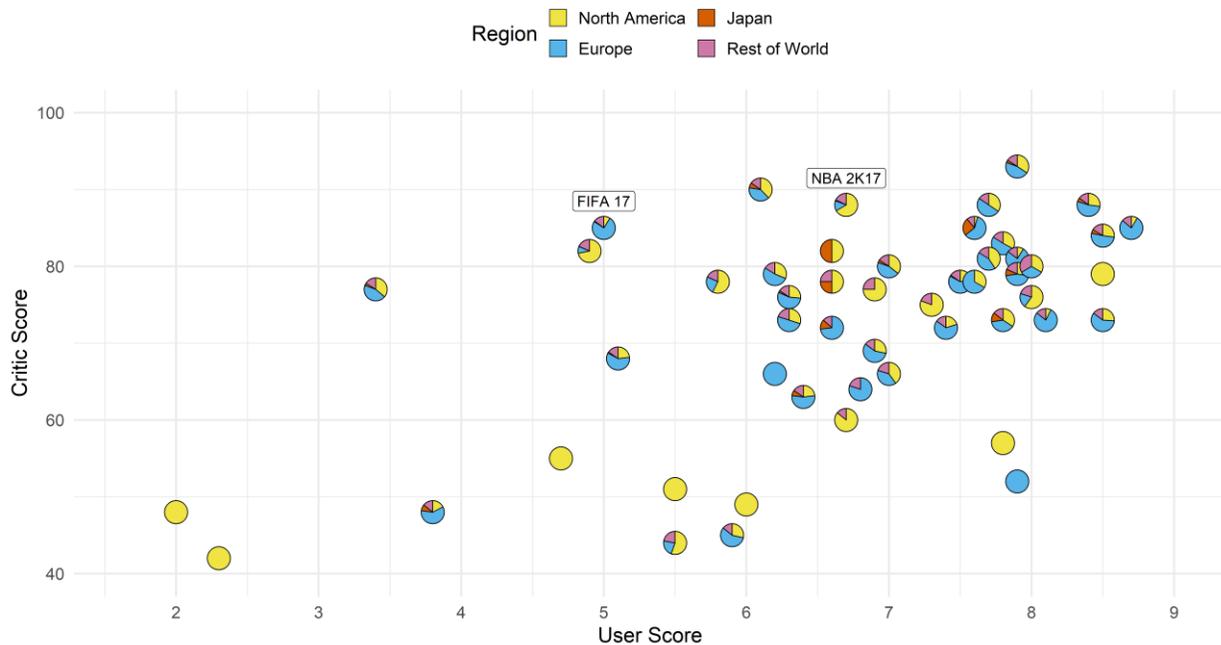

Figure 1. Pie-glyph scatterplot of critic versus user scores for selected games released in 2016. The data for this plot comes from Kirubi, 2017. The pie-glyphs show the proportions of game unit sales across the four regions North America, Europe, Japan, and the rest of the world. Two observations are labelled to highlight the distribution of sales for specific games.

Cleveland and McGill, 1985 demonstrated that visualisations with length based visual encodings (such as bar-charts) are generally superior to angle based encodings (pie-charts). However, in the 2d plot setting such as a scatterplot, using pie-charts overlaid on each point offers the following benefits over bar-charts.

(i) The circular shape of the pie-glyphs centred at the x-y location conveys additional visual information while maintaining the simplicity of a scatterplot.
(ii) A pie-glyph scatterplot uses polar coordinates for the pie within a cartesian coordinate system for the plot. Replacing the pie-glyph with a bar-chart glyph means the nested coordinate system is also cartesian, which is potentially a source of confusion when viewing the plot.

Other packages in R including scatterpie (Yu, 2024) and ggforce (Pedersen, 2022) offer functionality for creating pie-glyphs. However, unlike with PieGlyph, the pie-glyphs created using these functions are linked to the axes of the plot. Thus, the shape of the pie-glyphs changes with the aspect ratio of the plot and stretching the plot in either direction results in the pie-glyphs getting squished into ellipses (see this vignette for an example). In PieGlyph we solve this issue by creating a separate nested coordinate system for the pie-glyphs within the main coordinate system of the plot, so the pie-charts are independent of the axes on the plot. A primitive solution for achieving this could be to create each pie-glyph as an independent image and superimpose all images on the plot by using the

`geom_image_glyph` function from `ggmulti` (Xu and Oldford, 2022). However, this would be inefficient from a storage and time perspective. `PieGlyph` creates axis-independent pie-glyphs as native grid objects (grobs) which can be seamlessly integrated with the `ggplot2` machinery without any additional overhead. Since the pie-glyphs have their separate nested coordinate system they will always be circles with fixed radii, even if the underlying coordinate system or aspect ratio of the plot changes. This characteristic is particularly useful when working with spatial data enabling users to overlay pie-glyphs on maps and alter the underlying map projections without distorting the pie-glyphs (see Figure 2 and this vignette for examples). Moreover, this technique can be generalised to create any glyph as independent of the axes of the plot.

A potential downside of pie-glyphs is that due to their larger size they are more prone to over-plotting compared to simple points. Nevertheless, solutions like jittering can help mitigate this issue. Another limitation of static pie-chart glyphs is that only the relative proportion of different attributes can be visualised and not their raw counts (where there are associated raw counts). `PieGlyph` accounts for this by providing the option to create interactive pie-glyphs which, when hovered over, show a tooltip highlighting the raw count and percentage of each attribute shown in the pie-glyph. Figure 2 shows an example of such a plot where interactive pie-glyphs are superimposed on the map of Europe to illustrate the breakdown of the ages of mothers at the time of their first birth across different countries in 1999 (Figure 2a) and 2017 (Figure 2b). It indicates that, women in Western Europe tend to have a higher age at the time of first birth, compared to women in Eastern Europe. Furthermore, Figure 2 also highlights the trend of an increased age at the time of first birth in 2017 as compared to 1999 across all countries. Hovering over a pie-glyph would show the raw counts and percentages of mothers in each age group in the respective country (illustrated for Iceland in 1999 here). This interactivity is incorporated in `PieGlyph` using the `ggiraph` (Gohel and Skintzos, 2024) package.

`PieGlyph` is designed to dovetail with all features and extensions offered by `ggplot2` and `ggiraph`, thereby allowing users to customise every component of the visualisation. As we have seen, it can be used to effectively present spatial and temporal data. We have prepared a series of vignettes (see vignettes) with additional examples, including for visualising the results of regression and classification models. `PieGlyph` is particularly useful when visualising and interpreting statistical models fit to compositional data as there is a natural sum to one constraint on the data variables. In addition to replacing points in a scatterplot, pie-glyphs can be added to (for example) the axis on a bar-chart, or at particular points on a ternary diagram to convey additional information about the relative proportions of the compositional variables (see examples in Moral et al., 2023; Finn et al., 2024; Grange et al., 2024).

## Acknowledgements

CB and RV were supported by the Science Foundation Ireland Frontiers for the Future programme, grant number 19/FFP/6888.

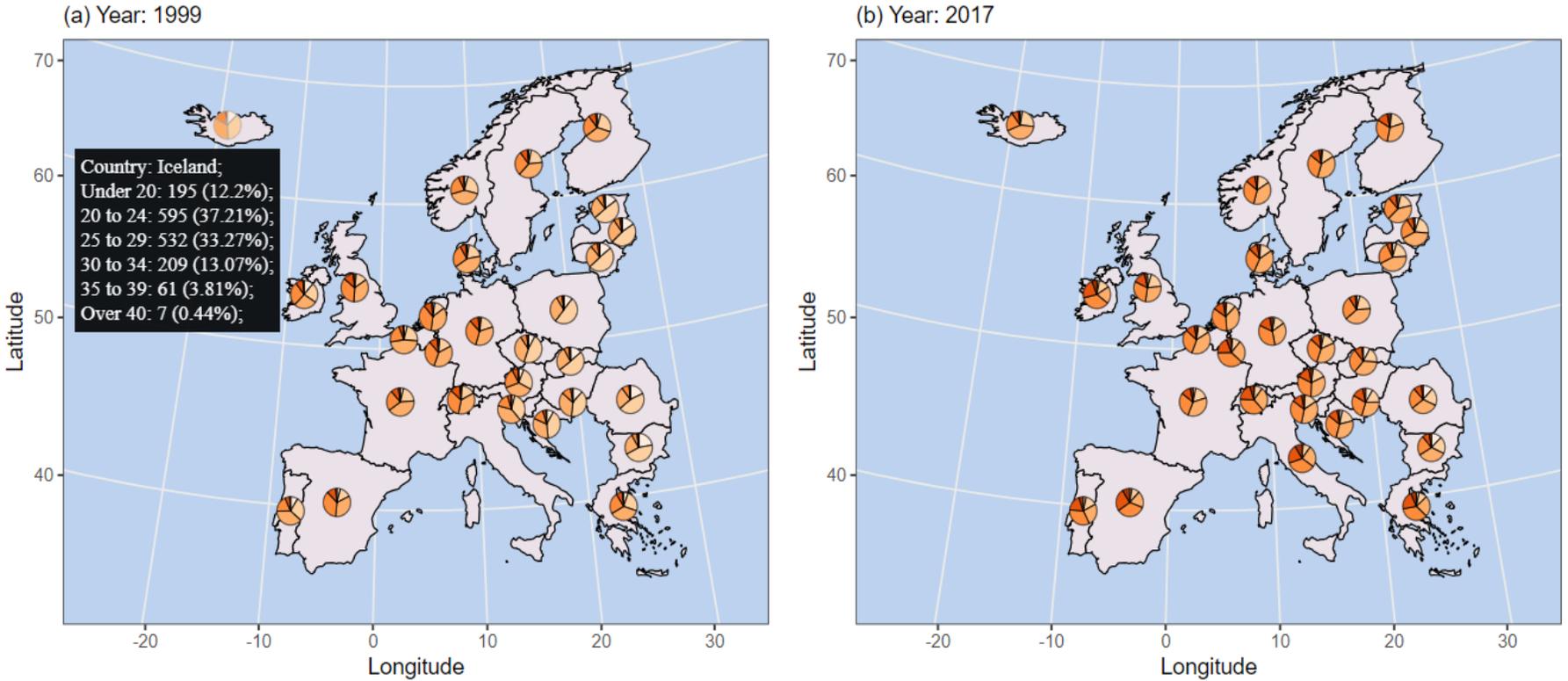

Figure 2. A map of Europe overlayed with pie-glyphs showing the proportion of mothers belonging to particular age group during their first birth in the year 1999 (a) and 2017 (b) in the respective countries. A tooltip is shown highlighting the raw counts in each age group in Iceland. The data for this plot is from Preda, 2021.